\documentstyle[12pt,psfig]{article}
\def\reference{\parskip 0pt\par\noindent\hangindent 0.5 truecm}

\begin{document}
%
%
\title{Galaxies Detected by the Dwingeloo Obscured Galaxies Survey}
%


\author{A. J. Rivers $^{1}$ \and
 P. A. Henning $^{1}$ \and
 R.C. Kraan-Korteweg $^{2}$ \and
} 

\date{}
\maketitle

{\center
$^1$ University of New Mexico, Dept. of Physics and Astronomy, 800 Yale NE, 
Albuquerque, USA, 87131\\andyr@as.unm.edu, henning@as.unm.edu\\[3mm]
$^2$ Dept. de Astronomia, Universidad de Guanajuato, Mexico\\kraan@astro.ugto.mx\\[3mm]
}

%
\begin{abstract}
The Dwingeloo Obscured Galaxies Survey (DOGS) is a 21-cm blind survey for 
galaxies hidden in the northern "Zone of Avoidance" (ZOA): the portion of
the optical extragalactic sky which is 
obscured by dust in the Milky Way.  Like the
Parkes southern hemisphere ZOA survey, the DOGS project is designed to reveal
hidden dynamically important nearby galaxies and to help "fill in the blanks"
in the local large scale structure.

To date, 36 galaxies have been detected by the Dwingeloo survey;
23 of these were previously unknown (no corresponding sources 
recorded in the NASA Extragalactic Database (NED)).  Among the interesting
detections are 3 nearby galaxies in the vicinity of NGC 6946 and 11 detections
in the Supergalactic plane crossing region.  VLA follow-up observations
have been conducted for several of the DOGS detections.
\end{abstract}


\bigskip

%
%

\section{Introduction}
Approximately 25\% of the optical extragalactic sky is obscured by the
dust and high stellar density of the Milky Way (figure 1).
Although diligent optical
and infrared searches for galaxies narrow this "Zone of Avoidance", in the
most heavily obscured regions near the Galactic plane, only radio surveys
consistently reveal hidden galaxies.  Radio galaxy searches near the Galactic
plane
can help to complete our knowledge of the nearby large scale structure and
explore the connectivity of superclusters and voids across the ZOA.  

A full survey of the northern ZOA has been recently completed using
the Dwingeloo 25-m radiotelescope and the Westerbork array in total 
power mode.  The survey region 
(30$^\circ$ $\leq  \ell  \leq$ 220$^\circ$ ; $|$b$|$ $\leq$ 5$^{\circ}\!\!$.25;
0 $\leq$ V$\mathrm{_{LSR}}$ $\leq$ 4000 km s$^{-1}$) was initially covered in a shallow
search by the 25-m 
(5 min per pointing; rms noise per channel $\sigma_{ch}$=175mJy)
for nearby, massive galaxies (Henning \emph{et al.} 1998).
Recently, a deeper survey sensitive to nearby dwarfs 
and to normal spirals at the survey limit
has been completed by the 25-m and the Westerbork array
in total power mode.
Results from the portion of the deep survey completed by the 25-m are reported
here.
Interesting regions covered by DOGS
include a portion of the Local Void, the
nearby IC342/Maffei group of galaxies and the region of sky where the Local
Supercluster crosses the Galactic plane.
 \begin{figure}[th]
 \centerline{\psfig{file=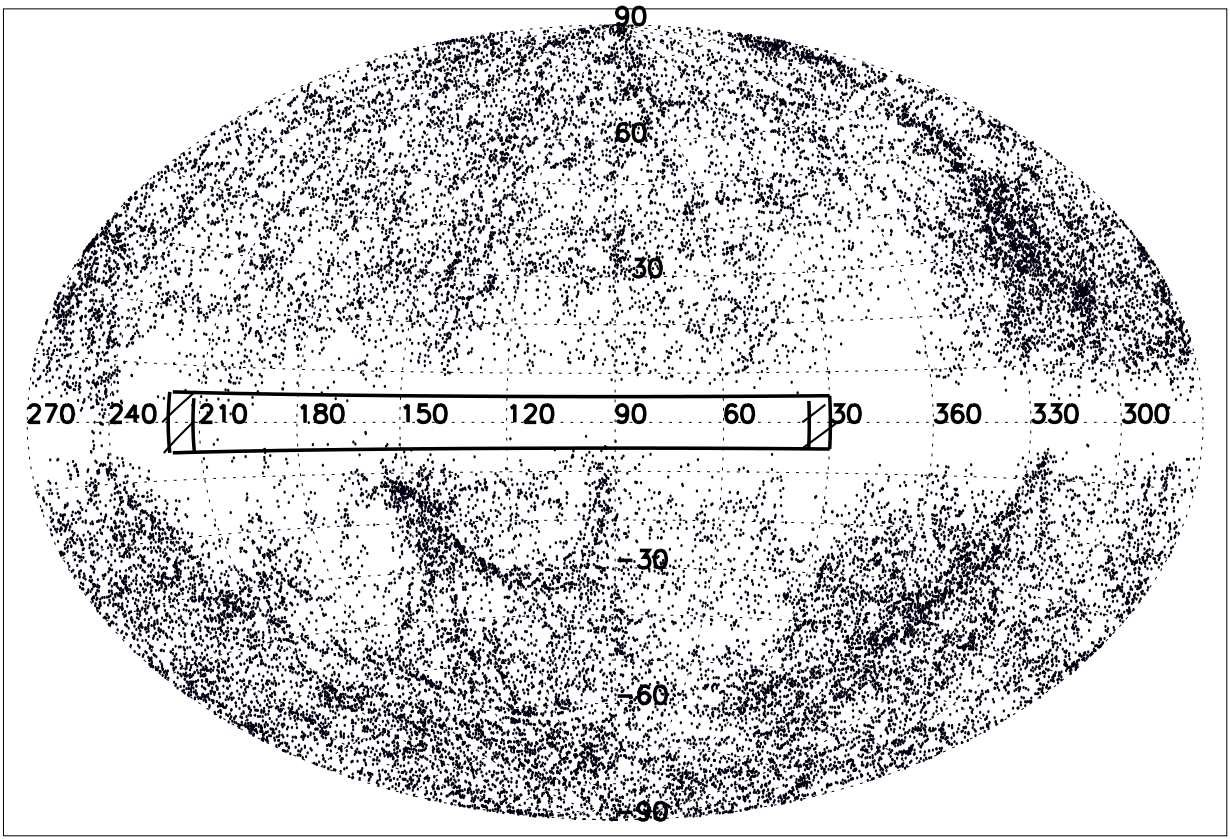,height=12cm}}
 \caption{A combined Galactic longitude vs. latitude 
	plot of approximately 30,000 galaxies collected
	from the MCG, UGC, and ESO optical catalogs shows the "Zone
	of Avoidance" caused by dust in the Galactic plane.  The spatial
	region covered by the Dwingeloo survey is indicated.  The
	hatched region represents the spatial overlap with the Parkes ZOA
	survey.}
 \end{figure}

\section{Telescope Parameters and Search Strategy}
The 25-m Dwingeloo telescope operating at 21-cm has a
half-power-beamwidth (HPBW)
of 0$^{\circ}\!\!$.6 which may be thought of as the survey
resolution.  A DAS-1000 channel autocorrelator spectrometer is utilized
in the telescope backend; the rms noise per channel is typically
$\sigma_{ch}$=40 mJy for a 1 hr integration.  
\begin{figure}[th]
\centerline{\psfig{file=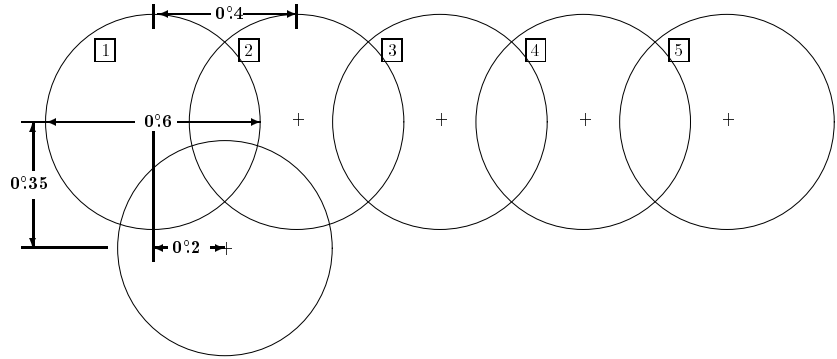,height=4.5cm}}
\caption{Distribution of survey grid points.  Galactic longitude spacing
is $\Delta \ell =0^{\circ}\!\!.4$ while the successive rows of constant
Galactic latitude are separated by $\Delta b=0^{\circ}\!\!.35$.  Each grid
point is marked with a + surrounded by a circle indicating the HPBW
of $0^{\circ}\!\!.6$.}
\end{figure}
Each DOGS observation consists of a sequence of 5 contiguous pointings
at constant Galactic latitude (figure 2).  Five On-Off pairs created from the
sequence ensure that a real galaxy will appear twice, once as a positive signal,
and again as a negative one, referenced against two independent scans.
Overlapping the constant latitude grids to form a honeycombed coverage of
the sky allows for detection of galaxies in adjacent pointings
and facilitates a more accurate determination of their positions.  A similar
observing strategy was employed during the Westerbork portion of the survey;
beam size and velocity coverage were identical and a similar survey 
sensitivity achieved.  A 15 pointing grid was used for efficiency reasons,
but the core On-Off pair strategy remained the same.
The 
complete survey incorporates approximately 15,000 partially overlapping
pointings.
\section{Results}
Approximately 60\% of the DOGS survey was completed using the Dwingeloo 25-m
telescope.  In this portion of the survey, 36 galaxies were confirmed, 23
of which were previously unknown (no NED counterpart).  The number of galaxies
detected is consistent with calculations based on an assumed HI mass function
(Zwaan \emph{et al.} 1997)
and the survey sensitivity which predict between 50 and 100 detections within
the survey range.  Recent Westerbork observations completed the survey and
galaxies discovered will soon be incorporated into the Dwingeloo sample 
(\emph{cf.} figure 3 for location of detected galaxies).

\begin{figure}[th]
\begin{center}
\centerline{\psfig{file=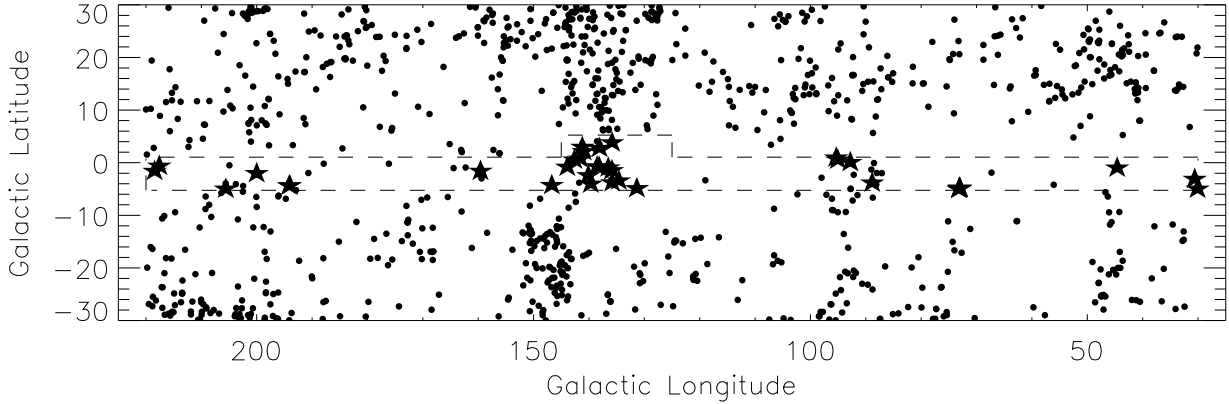,height=5cm}}
\caption{Spatial distribution of DOGS sources (indicated by $\star$)
combined with
Lyon-Meudon Extragalactic Database (LEDA)
galaxies out to V$\mathrm{_{LSR}} \leq$ 4000 km s$^{-1}$ ($\bullet$).
The 60\% of the survey covered by the 25-m and analyzed in this paper is shown 
by the dashed line.}
\end{center}
\end{figure}

Five of the 36 sources were originally identified
by the shallow survey 
including Dwingeloo 1 and Maffei 2, both members
of the nearby Maffei / IC 342 group of galaxies
(Kraan-Korteweg \emph{et al.} 1994).  During the deep survey another known 
group member, MB 1 (McCall \& Buta 1995), was identified 
and two additional members still await
confirmation observations.

The most significant nearby, previously unknown galaxy identified by DOGS was
Dwingeloo 1.  Given the 80\% coverage of the survey region by the shallow
survey (Henning \emph{et al.} 1998), 
chances are low that a massive nearby spiral
was missed,
since nearby galaxies appear in many adjacent pointings, all of which would
have to be missed for the galaxy to escape detection.  Thus, it is
fairly unlikely that there is another previously unidentified massive galaxy
whose gravitational influence significantly impacts Local Group peculiar motion
or internal dynamics in the area covered by the survey.

\subsection{The Supergalactic Plane Crossing Region and the Local Void}
Although the Dwingeloo survey is limited in survey depth (V$\mathrm{_{LSR}}\leq$ 4000 km
s$^{-1}$), two primary large scale structures fall partially 
within the survey range: the Local Void and the Local Supercluster.
Eleven galaxies were discovered
in the survey coverage of the Local Supercluster crossing region 
($\ell\sim$140$^\circ$; $|$b$|$ $\leq$ 5$^{\circ}\!\!$.25);
6 of these sources are
noted in NED.  Known structures
appear continuous and well defined across the Galactic plane
with a narrow bridge of galaxies visible at $\ell\sim$142$^\circ$
and V$\mathrm{_{LSR}}$ $\sim$ 1400 km s$^{-1}$ .

Near the Local Void ($\ell\sim$33$^\circ$, b$\sim$-15$^\circ$), 
Marzke \emph{et al.} (1996) and Roman \emph{et al.} (1998)
found evidence for a nearby cluster at $\sim$1500 km s$^{-1}$.  Two previously 
unknown galaxies were detected in this region, adding support for the
hypothesized overdensity.  These galaxies, Dw030.6-2.4 ($\ell$=30.60, b=-2.48,
V$\mathrm{_{LSR}}$=1480 km s$^{-1}$) and Dw030.1-4.3 ($\ell$=30.09, b=-4.35,
V$\mathrm{_{LSR}}$=1528 km s$^{-1}$), were independently identified by the Parkes ZOA
survey (Henning \emph{et al.}, this volume).

\subsection{NGC 6946 Group}
Three dwarf galaxies were detected in the region of NGC 6946 ($\ell$=95.72;
b=11.67) suggesting the possibility of a new nearby group.  Of these,
Dw095.0+1.0 was originally recorded as a 
compact High Velocity Cloud (HVC) (Wakker 1990),
but present data suggest it is in fact a nearby dwarf galaxy.  With a 
velocity of V$_{GSR}$=368 km s$^{-1}$, it has the highest redshift of any HVC
in the Wakker catalog and the 50\% velocity width of $\Delta$V$_{50}$=100
km s$^{-1}$ seen in the Dwingeloo spectrum is significantly broader than
the 20-30 km s$^{-1}$ velocity dispersion generally observed in HVCs.
Naively assuming this galaxy is at the same distance as NGC 6946
($\sim$6 Mpc, Sharina \emph{et al.} 1997) yields M$\mathrm{_{HI}}\simeq$
4$\times$10$^8$ M$_\odot$.
\begin{figure}[th]
\psfig{file=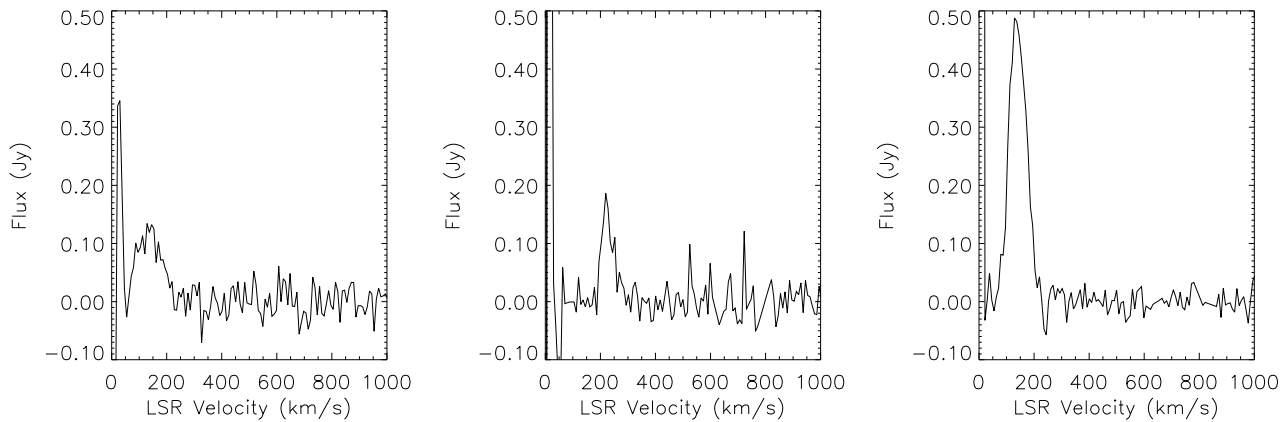,height=5.5cm}
\caption{Spectra from the Dwingeloo 25-m telescope of three galaxies found
        in the vicinity of NGC 6946.  Pictured are, from left to right,
	IRAS 21189+4503, Dw092.8+0.5 and Dw095.0+1.0}
\end{figure}
Also detected in the area were IRAS 21189+4503 (Nakanishi 1997) and
Dw092.8+0.5.  Assuming
a 6 Mpc distance yields M$_{HI}\simeq$ 
2$\times$10$^8$ M$_\odot$ and 1$\times$10$^8$ M$_\odot$
respectively.  The estimated neutral hydrogen masses are typical of dwarf
galaxies, consistent with the gaussian shape of the 21-cm profiles (figure 4). 
With the recent discovery of the LSB galaxy Cepheus 1, (Burton \emph{et al.},
\emph{in press}) 7 galaxies with recessional velocites V$\mathrm{_{LSR}} \leq$
250 km s$^{-1}$ have been identified within 15$^\circ$ of NGC 6946 (table 1).
If these galaxies do in fact 
signify a new nearby group, this group would lie some 40$^\circ$ from the
Supergalactic plane, considerably more than any other known group in the 
local universe.  The potential exists for the discovery of additional galaxies
in this group since the data recently collected by the Westerbork array 
for +1$^{\circ}\!\!$.05 $\leq$ b $\leq$
+5$^{\circ}\!\!$.25 has not yet been analyzed. 

  \begin{table}[th]
\begin{center}
    \begin{tabular}{|l|c|c|c|c|c|}
    \hline
    \multicolumn{1}{|c}{Galaxy Name} &
    \multicolumn{1}{|c}{$\ell$} &
    \multicolumn{1}{|c}{b} & \multicolumn{1}{|c|}{V$\mathrm{_{LSR}}$} &
    \multicolumn{1}{|c}{log(M$\mathrm{_{HI}}$)} &
    \multicolumn{1}{|c|}{Reference} \\
    \multicolumn{1}{|c}{} & \multicolumn{1}{|c}{} &
    \multicolumn{1}{|c}{} &
    \multicolumn{1}{|c|}{km s$^{-1}$} &
    \multicolumn{1}{|c}{M$_\odot$} &
    \multicolumn{1}{|c|}{} \\
    \hline
	HKK L150 & 96.03 & 12.35 & 132 & 7.1& Huchtmeier \emph{et al.} 1997 \\
	UGC 11583  & 95.63 & 12.31 & 127 & 8.2& Huchtmeier \emph{et al.} 1997 \\
	HKK L149 & 95.57 & 12.22 & 126 & 8.1 & Huchtmeier \emph{et al.} 1997 \\
	NGC 6946 & 95.72 & 11.67 & 48 & 10.1 & LEDA\\
	Cepheus 1 & 94.38&8.01 & 65&9.1&Burton \emph{et al.} \emph{in press} \\
	Dw095.0+1.0 & 95.05 & 1.16 & 142 & 8.6:&Henning \emph{et al.} 1998 \\
	Dw092.8+0.5 & 92.82 & 0.53 & 230 & 7.8:& \\
	IRAS 21189+4503 & 88.85 & -3.20 & 228 & 8.3: &Nakanishi \emph{et al.} 1997 \\
    \hline
    \end{tabular}
    \caption{Possible members of the NGC 6946 group.  HI masses are calculated
	based on the assumption that 
	all galaxies lie at the same distance as NGC 6946 (6 Mpc,
	Sharina \emph{et al.} 1997).  The HI mass of NGC 6946 is based on
	the mean of the 6 flux entries in LEDA which are corrected
	for the beam-filling effect.  The uncertainty
	in the masses of the 
	three Dwingeloo detected sources is primarily due to the uncertainty
	in the positions of the galaxies. }

\end{center}
  \end{table}
\subsection{Results from Synthesis Follow-ups of DOGS Sources}
Synthesis observations of Dwingeloo galaxies were conducted with the VLA and
WSRT in 1997.  Snapshot WSRT observations yielded
positions of DOGS sources though sensitivities were not adequate for
detailed mapping.  VLA follow-ups yielded a few interesting results.
\begin{figure}[th]
\centerline{\psfig{file=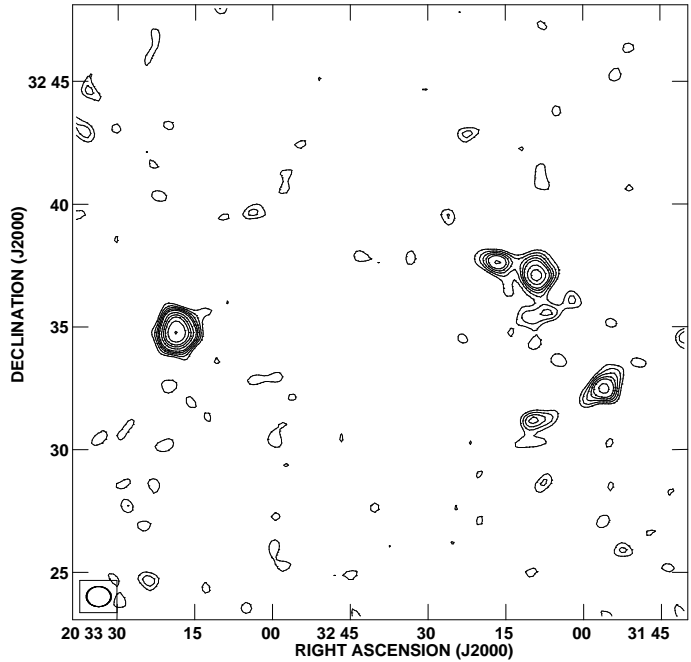,height=9cm}}
\caption{A synthesis map of DOGS detection Dw073.2-4.2 from the VLA.
A galaxy group is 
revealed in a sample velocity slice (V$\mathrm{_{LSR}}$=3219 km s$^{-1}$, 
$\Delta$v $\simeq$ 10 km s$^{-1}$).}
\end{figure}
A galaxy group unresolved by the Dwingeloo $0^{\circ}\!\!.6$
beam was resolved into 5 galaxies by follow-up VLA observations (figure 5).
Since early synthesis follow-ups to Parkes ZOA detections also have revealed 
interacting galaxy groups (Staveley-Smith \emph{et al.} 1998), it will
be interesting to see if this is commonly seen in HI selected samples.
The total HI mass measured for this group was 2$\times$10$^9$
M$_\odot$h$_0^{-2}$ with a mean velocity
of V$\mathrm{_{LSR}}$=3222 km s$^{-1}$ and a calculated group diameter of 
approximately 150 h$_0^{-1}$ kpc.  No optical counterparts were identified on
POSS E plates under significant
extinction (A$_B \simeq$ 4.5 mag from DIRBE reddening maps,
Schlegel \emph{et al.} 1998).
A previously unknown low mass 
companion to IRAS 05596+1451 was also discovered in a separate VLA synthesis
map.  
\section{Conclusions and Future Directions}

A wide field flux limited survey 
for HI luminous galaxies hidden in the northern
ZOA has been completed.  With approximately 60\% of the
data analyzed we have detected 36 galaxies, 23 of which were previously
unknown.  Given the number found to date, 
in the full survey we expect to detect around 60 sources, which falls within
our survey predictions of 50-100 galaxies.  

Since the DOGS slice covers a
wide range of the extragalactic 
sky including portions of the Local Supercluster and the
Local Void, the 
collected sample will be well suited for a comparison study 
of galaxies in different environments.

Efforts are also underway to quantify the Galactic extinction, a poorly known 
quantity near the Galactic plane.  Although the recently released
DIRBE reddening maps (Schlegel \emph{et al.} 1998) 
prove useful, they remain uncalibrated at low latitude
($|$b$|$ $\leq$ 10$^{\circ}$).  We are exploring the possibility of using
near-infrared colors of galaxies from 2MASS data to independently determine
the extinction and to thereby determine calibration constants and uncertainties
for the DIRBE maps at low latitude.
\section*{Acknowledgements}
We are grateful to
A. Foley and D. Moorrees for their supervision of the
25-m day-to-day telescope operations,
to T. Hess for the development of data translation routines, and to 
O. Lahav and W. B. Burton for helpful discussions. 
We would also like to thank
J.M. van der Hulst and R. Braun for their help in instigating and supporting
the Westerbork portion of the DOGS project.
The Dwingeloo 25-m radio telescope is
supported by the Netherlands Foundation for Scientific Research (NWO).
This research has made
use of the NASA/IPAC Extragalactic Database (NED) and the Lyon-Meudon
Extragalactic Database (LEDA).
The research of P.A. Henning is supported by  NSF Faculty Early Career
Development (CAREER) Program award AST 95-02268.

\section*{References}






\reference Burton, W.B., Braun, R., Walterbos, R.A.M., Hoopes, C.G. AJ,
\emph{in press}
\reference Henning, P.A., Kraan-Korteweg, R.C., Rivers, A.J., Loan, A.J,
Lahav, O., Burton, W.B. 1998, AJ, 115, 584
\reference Huchtmeier, W.K., Karachentsev, I.D., Karachentseva, V.E. 1997,
A\&A, 322, 375
\reference Kraan-Korteweg, R.C., Loan, A.J., Burton, W.B., Lahav, O.,
Ferguson, H.C., Henning, P.A., \& Lynden-Bell, D. 1994, Nature, 372, 77
\reference Marzke, R.O., Huchra, J.P., \& Geller, M.J. 1996, AJ, 112, 1803
\reference McCall, M.L, \& Buta, R.J., 1995, AJ, 109, 2460
\reference Nakanishi, K., Takata, T., Yamada, T., Takeuchi, T., Shiroya, R., 
Miyazawa, M., Watanabe, S., \& Saito, M. 1997, ApJS, 112, 245
\reference Roman, A.T., Takeuchi, T., Nakanishi, K., \& Saito, M. 1998, PASJ,
50, 47
\reference Schlegel, D.J., Finkbeiner, D.P., Davis, M. 1998, ApJ, 500, 525
\reference Sharina, M.E., Karachentsev, I.D., \& Tikhonov, N.A. 1997, 
Astr. Letters, 23, 373
\reference Staveley-Smith, L., Juraszek, S., Koribalski, B.S., Ekers, R.D., 
Green, A.J., Haynes, R.F., Henning, P.A., Kesteven, M.J., Kraan-Korteweg,
R.C., Price, R.M., Sadler, E.M. 1998, AJ, 116, 2717
\reference Wakker, B.P., 1990, Ph.D. thesis, Groningen Univ.
\reference Zwaan, M., Briggs, F., \& Sprayberry, D. 1997, PASA, 14, 126

\end{document}